\documentclass[fleqn,twoside]{article}
\usepackage[headings]{espcrc2}

\usepackage{amssymb}

\readRCS
$Id: espcrc2.tex,v 1.2 2004/02/24 11:22:11 spepping Exp $
\usepackage{graphicx}
\usepackage[figuresright]{rotating}

\newcommand{\AmS}{{\protect\the\textfont2
  A\kern-.1667em\lower.5ex\hbox{M}\kern-.125emS}}

\def\fig#1{fig.~{\ref{#1}}}

\def\tab#1{table~{\ref{#1}}}

\def\spa#1.#2{\left\langle#1\,#2\right\rangle}
\def\spb#1.#2{\left[#1\,#2\right]}
\def\tree{{\rm tree}}
\def\Loop{{\rm loop}}

\def\eps{\epsilon}

\def\nn{\nonumber}

\def\eqn#1{eq.~(\ref{#1})}

\def\NeqFour{{{\cal N}=4}}

\def\NeqEight{{{\cal N}=8}}

\def\be{\begin{equation}}
\def\ee{\end{equation}}
\def\bea{\begin{eqnarray}}
\def\eea{\end{eqnarray}}
\def\ba{\begin{eqnarray}}
\def\ea{\end{eqnarray}}

\def\tree{{\rm tree}}

\newbox\charbox
\newbox\slabox
\def\s#1{{      
        \setbox\charbox=\hbox{$#1$}
        \setbox\slabox=\hbox{$/$}
        \dimen\charbox=\ht\slabox
        \advance\dimen\charbox by -\dp\slabox
        \advance\dimen\charbox by -\ht\charbox
        \advance\dimen\charbox by \dp\charbox
        \divide\dimen\charbox by 2
        \raise-\dimen\charbox\hbox to \wd\charbox{\hss/\hss}
        \llap{$#1$} }}

\def\n{{\tilde n}}
\def\f{\widetilde f}
\def\tree{{\rm tree}}

\def\nn{\nonumber}

\newif\ifdraft
\drafttrue
\newif\ifpreprint
\preprinttrue

\def\fig#1{fig.~{\ref{#1}}}

\def\eqn#1{eq.~(\ref{#1})}

\def\eq<ns#1#2{eqs.~(\ref{#1}) and~(\ref{#2})}

\def\tab#1{table~{\ref{#1}}}

\def\mud{\lambda}

\title{
\vskip - 1 cm 
\hbox{\normalsize
UCLA/10/TEP/104$\null\hskip 9.1 cm \null$ 
Saclay--IPhT--T10/086 $\hskip 0.55cm \null$}
$\null$ \hskip 15 cm $\null$
The Structure of Multiloop Amplitudes in Gauge and Gravity Theories}

\author{
Z.~Bern\address[UCLA]{Department of Physics and Astronomy, UCLA, 
   Los Angeles, CA 90095-1547, USA}\thanks{Presenter at Loops and Legs in 
  Quantum Field Theory, W\"orlitz, Germany, April 25-30, 2010},
J.~J.~M. Carrasco\addressmark[UCLA] and
H.~Johansson\address{Institut de Physique Th\'eorique, CEA--Saclay,
          F--91191 Gif-sur-Yvette cedex, France}
}
       
\runtitle{ Multiloop Amplitudes in Gauge and Gravity Theories}
\runauthor{Z.~Bern, J.~J.~M. Carrasco and H.~Johansson}

\begin{document}


\begin{abstract}
We review the recently discovered duality between color and kinematics
in gauge theories.  This duality leads to a remarkably simple
double-copy relation between diagrammatic numerators of gravity
scattering amplitudes and gauge-theory ones.  We summarize nontrivial
evidence that the duality and double-copy property holds to all loop
orders.  We also comment on other developments, including a proof that
the gauge-theory duality leads to the gravity double-copy property,
and the identification of gauge-theory Lagrangians whose double copies
yield gravity Lagrangians.
\end{abstract}


\maketitle

\section{Introduction}

Gauge theory and gravity scattering amplitudes have a far richer
structure than evident from their respective Lagrangians.  As one such
example, in this talk we will describe a recently discovered duality
between color and kinematic numerators of gauge-theory scattering
amplitude diagrams~\cite{BCJ,BCJLoops}.  Remarkably, this duality
appears to have important implications for gravity: when the
gauge-theory numerators satisfy the duality, the numerators of
corresponding gravity theories are given by a double copy of the gauge
theory numerators, diagram by diagram~\cite{BCJ}, as demonstrated
recently~\cite{Square}.  The double-copy property has the benefit of
greatly clarifying the mysterious Kawai, Lewellen and Tye (KLT)
relation between gauge and gravity tree amplitudes~\cite{KLT}.  In
this talk we will focus on the recent progress in extending the
duality and double-copy properties to loop level~\cite{BCJLoops}. We
also summarize the structure of gauge-theory Lagrangians whose Feynman
diagrams satisfy the duality, leading to gravity Lagrangians that
exhibit the double-copy property~\cite{Square}.

\begin{figure}[tbh]
\begin{center}
\includegraphics[clip,scale=0.32]{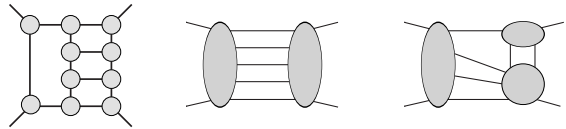}
\end{center}
\vskip -1.2 cm
\caption{\small Some examples of generalized cuts at four loops
which decompose loop amplitudes into sums of products of tree amplitudes.}
\label{CutExamplesFigure}
\vskip -.3 cm 
\end{figure}

Many of the recent developments for scattering amplitudes rely on
on-shell methods, which include on-shell recursion~\cite{BCFW} at tree
level and the unitarity method~\cite{UnitarityMethod} at loop level. A
particularly powerful approach to generalized unitarity cut
constructions~\cite{GeneralizedUnitarity} is the method of maximal
cuts~\cite{FiveLoop,ManifestThreeLoop}, organizing the
calculation starting from unitarity cuts where all propagators are cut
and then systematically reducing the number of cut propagators.
On-shell methods allow one to construct new amplitudes using simpler
on-shell amplitudes as input. Cuts that decompose loop amplitudes into
tree amplitudes, such as the sample four-loop ones displayed in
\fig{CutExamplesFigure}, are generally the most advantageous ones to
use.  The unitarity method gives a set of rules for reconstructing
complete amplitudes from cuts. As discussed in a variety of talks at
this conference~\cite{ConfTalks}, the unitarity method is by now a
standard tool in loop computations.  As one state-of-the-art example
from collider physics, it plays a central role in progress towards the
long-awaited NLO calculation of $W+4$-jet production at the LHC,
described in this conference~\cite{MaitreTalk}.

\begin{figure}[tb]
\begin{center}
\includegraphics[clip,scale=0.32]{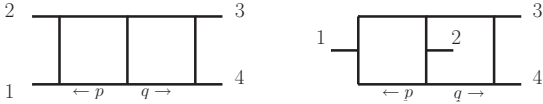}
\end{center}
\vskip -1. cm
\caption{\small The two-loop QCD amplitudes identical helicity 
amplitudes can be expressed in terms of planar and non-planar
double-box integrals, as well as a ``bow-tie'' integral (not displayed
here). 
\vskip -.2 cm 
}
\label{TwoLoopIntegralFigure}
\end{figure}

Multiloop calculations are typically rather involved, yielding
expressions with no easily accessible structure or pattern.  Consider,
{\it e.g.,} the rather lengthy results for $gg \rightarrow gg$
scattering amplitudes in QCD, given in refs.~\cite{ggggTwoloop}.
Might there be a structure hidden in these rather opaque expressions?
To answer this question, it is easiest to begin by looking at one the
simplest $2 \rightarrow 2$ two-loop amplitudes---identical helicity
scattering in QCD---first computed in ref.~\cite{TwoLoopAllPlus}.

The all-plus helicity amplitude is given in terms of three integrals,
two of which are displayed in \fig{TwoLoopIntegralFigure}.  The planar
and non-planar double-box contributions shown in
\fig{TwoLoopIntegralFigure} may be expressed in terms of loop
integrals as in ref.~\cite{TwoLoopAllPlus},
\begin{eqnarray}
A^{\rm P}_{1234} \hskip -.2 cm & =& \hskip -.2 cm 
 i \,{\spb1.2\spb3.4 \over \spa1.2\spa3.4 }\, s_{12}
  \, I_4^{\rm P} \Bigl[ N(p,q)\Bigr] \,, \nn \\
A^{\rm NP}_{12;34} \hskip -.2 cm &= & \hskip -.2 cm
 i \, { \spb1.2\spb3.4 \over \spa1.2\spa3.4 } \, s_{12} \, 
I_4^{\rm NP} \Bigl[ N(p,q) \Bigr]\,, \nn
\end{eqnarray}
where the  $N(p,q)$ represent numerator polynomials depending
on the loop momenta $p$ and $q$.
Remarkably, the polynomials (and the prefactors) for both 
integrals are identical.  The shared numerator is given by, 
\begin{eqnarray}
&&  \hskip -.7 cm 
N(p,q) = (D_s-2) ( \mud_p^2 \, \mud_q^2 
                   + \mud_p^2 \, \mud_{p+q}^2 
                   + \mud_q^2 \, \mud_{p+q}^2 ) \nn\\
&& \null \hskip 1 cm 
+ 16 \Bigl( (\mud_p \cdot \mud_q)^2
           - \mud_p^2 \, \mud_q^2 \Bigr)\,, \nn
\end{eqnarray}
where the vectors $\vec\mud_i$ represent the $(-2\eps)$-dimensional
components of the loop momenta $\ell_i$; that is, $\ell_i \equiv
\ell_i^{[4]} + \mud_i$, where $\ell_i^{[4]}$ has only four-dimensional
components and we take  $(-2\eps)$ to be positive. 
The propagators of the integrals are of course different
and correspond to the two diagrams in \fig{TwoLoopIntegralFigure}.
The prefactor depends on spinor inner products $\spa{i}.{j}$ and 
$\spb{i}.{j}$;
further details on spinor-helicity notation may be found in, for example,
ref.~\cite{TreeReview}. We denote the number of gluon states
circulating in the loop by $D_s - 2$, where $D_s$ is the dimension of
spacetime.  ($D_s$ is the dimension of the polarization vectors
which can be taken to be independent of the dimension of the loop momenta.)

The above result has a rather striking feature: The numerators of the
planar and non-planar integrals are identical. This feature is rather
obscure from a Feynman diagram point of view, where no discernible
relationship between the planar and non-planar contributions is
apparent.  This curious feature seems to be an important clue for novel
structures in gauge-theory amplitudes.  But what might it be?  We know
now~\cite{BCJ,BCJLoops} that this curiosity is a hint of hidden
structures, including:
\begin{itemize}

\item A new duality between color and kinematics.

\item All-loop relations between planar and non-planar diagrammatic integrands.

\item A double-copy structure of gravity diagram kinematic numerators
  in terms of gauge-theory ones.

\end{itemize}

Although these properties are not yet completely proven, below we
discuss evidence that these properties hold to all loop
orders.


\section{A duality between color and kinematics}

Any amplitude with all particles in the adjoint representation of
the gauge group can be arranged into the form, 
\begin{equation}
{\cal A}^\tree_n(1,2,3,\ldots,n)=\sum_{i}
                 \frac{n_i \, c_i}{\prod_{\alpha_i} p^2_{\alpha_i}}\,,
\label{AGauge}
\end{equation}
where the sum runs over the set of $n$-point diagrams with only cubic
vertices and we suppress factors of the coupling constant.  The
product runs over all propagators (internal lines) $1/p^2_{\alpha_i}$
of each diagram.  The $c_i$ are the color factors obtained by dressing
every three vertex with an $\f^{abc} = i \sqrt{2} f^{abc}$ structure
constant, and the $n_i$ are kinematic numerator factors depending on
momenta, polarizations and spinors (and Grassmann parameters for
supersymmetric amplitudes expressed in superspace).  The form
(\ref{AGauge}) follows trivially from Feynman diagrams, by
representing all contact terms as inverse propagators in the kinematic
numerators, which then cancel propagators.

\subsection{Tree-level duality}
According to the duality proposal of ref.~\cite{BCJ}, arrangements of
the diagrammatic numerators in \eqn{AGauge} exist such that they
satisfy equations in one-to-one correspondence with the color Jacobi
identities.  Specifically, we demand that every color
Jacobi identity induces a kinematic identity:
\begin{equation}
c_i = c_j-c_k \;  \Rightarrow \;  n_i = n_j-n_k \,.
\label{BCJDuality}
\end{equation}
This duality is expected to hold in a large variety of theories,
including supersymmetric extensions of Yang-Mills theory.  At four
points the duality is automatically satisfied for any choice of valid
numerators~\cite{Halzen,BCJ}, but at higher points the existence of such
arrangements is rather nontrivial.  Indeed there is, as of yet, no complete
proof that this can always be accomplished.  Surprisingly, this
duality implies non-trivial relations between the color-ordered
partial amplitudes of gauge theory~\cite{BCJ}.  A proof of these
amplitude relations has recently been given both in field theory and
in string theory~\cite{Bjerrum1,Feng}.

\subsection{Gravity as a double copy of gauge theory}

Remarkably, once the gauge-theory amplitudes are arranged into a form
satisfying the duality (\ref{BCJDuality}), gravity tree amplitudes are
given by a double copy of gauge-theory numerator factors~\cite{BCJ},
\begin{equation}
-i {\cal M}^\tree_n(1,2,\ldots,n)=
    \sum_{i}{ \frac{n_i\, \n_i}{\prod_{\alpha_i}{p^2_{\alpha_i}}}} \,,
\label{Squaring}
\end{equation}
where the $\n_i$ represent numerator factors of a second gauge theory
amplitude, the sum runs over the same set of diagrams as in
\eqn{AGauge}, and we suppressed the gravitational coupling constant.
This formula recently has been proven for pure gravity and for
$\NeqEight$ supergravity, under the assumption that the
duality~(\ref{BCJDuality}) holds in the corresponding gauge theories,
by use of on-shell recursion~\cite{Square}.  If pure-Yang-Mills
amplitudes are used, the obtained gravity amplitudes correspond to
Einstein gravity coupled to an antisymmetric tensor and dilaton; the
$n$-graviton tree-level amplitudes of this theory are those of pure
Einstein gravity.  If both families of kinematic factors are for the
$\NeqFour$ super-Yang-Mills theory, the gravity theory amplitudes are
those of $\NeqEight$ supergravity.  For this double-copy relation to
hold, at least one family of numerators ($n_i$ or $\n_i$) needs to
satisfy the
duality~(\ref{BCJDuality})~\cite{BCJ,Square}. Additionally, the
$\tilde{n}_i$ numerator factors need not come from the same theory as
the $n_i$ factors.  This allows for the construction of gravity
amplitudes with varying amounts of supersymmetry.  The duality has
also been studied in string theory~\cite{Mafra}.  We note that the
heterotic string offers a natural venue for
understanding these properties because of the parallel treatment of
color and kinematics~\cite{Tye}.

\subsection{Lagrangian insight}

If one compares the Yang-Mills and Einstein gravity Lagrangians
\begin{equation}
{\cal L}^{\rm YM} = {1\over 4 g^2} F^2 \,, \hskip 1 cm 
{\cal L}^{\rm G} = {2\over \kappa^2} \sqrt{-g} R \,,
\label{Lagrangians}
\end{equation}
it seems rather puzzling that gravity can be a double copy of gauge
theory.  Indeed, the perturbative expansions appear
to be completely different: expanding the metric around flat space, we
find that gravity is composed of an infinite sequence of increasingly
complicated vertices, whereas the Yang-Mills Lagrangian has 
local interactions that terminate at four points.  On the other hand, the
KLT relations map the tree-level scattering amplitudes of gravity
to those of gauge theory.  This suggests that the standard
Lagrangians~(\ref{Lagrangians}) obscure the relation between the two.

Some initial steps in understanding the relationship between gravity
and gauge theory at the level of the Lagrangian were carried out in
ref.~\cite{BernGrant}. In particular, that paper developed 
some tricks for separating the
graviton Lorentz indices into ``left'' and ``right'' classes,
consistent with the double-copy property.  (See also
ref.~\cite{Siegel}.)  However, the connection to gauge theory remained
rather mysterious.  As realized in ref.~\cite{Square}, the key missing 
ingredient was the color-kinematic duality of ref.~\cite{BCJ}.

To see how to make the proper rearrangement, consider first the
Yang-Mills Lagrangian.  The initial step is to rearrange the Lagrangian
so its Feynman diagrams satisfy the color-kinematic duality.  As noted
earlier for four-point amplitudes, the duality is automatic---the
ordinary form of the Yang-Mills Lagrangian generates diagrams that
satisfy the duality. Beyond four points, for the duality to be
satisfied directly by the Feynman diagrams, we must modify the
Lagrangian yet leave the amplitudes unchanged.  For five points, this
may be accomplished by adding the following term to the Yang-Mills
Lagrangian~\cite{Square}:
\begin{eqnarray}
 \mathcal{L}'_5 \hskip -.2 cm &=& \hskip -.2 cm 
 -\frac{1}{2} g^3 (f^{a_1a_2b}f^{ba_3c} \nn \\
&& \hskip 1cm 
+ f^{a_2a_3b}f^{ba_1c}
   + f^{a_3a_1b}f^{ba_2c})f^{ca_4a_5} \nonumber \\
  && \quad
   \times \partial^{\phantom{a_1}}_{[\mu} A^{a_1}_{\nu]} A^{a_2}_\rho A^{a_3\mu}
  \frac{1}{\Box}(A^{a_4\nu} A^{a_5\rho}) \,.
\label{Lagragian5Alt}
\end{eqnarray}
This new term leaves the amplitudes unchanged because it vanishes
identically by the color-Jacobi identity. Nevertheless, the canceling
terms have different color structures and thus are associated with
different diagrams. These terms modify the individual diagrams so that
the duality holds.  The additional term (\ref{Lagragian5Alt}) is,
however, nonlocal.  We can make the Lagrangian local by introducing a
set of auxiliary fields as explained in ref.~\cite{Square}.  In fact,
with a sufficient number of such fields, all higher-point interactions
can be replaced by three-point interactions.  Once the gauge-theory
Lagrangian has been put into this form, a gravity Lagrangian which
yields the correct amplitudes is given simply by two copies of the
gauge theory Lagrangian, as described in ref.~\cite{Square}, with the
identification $A^a_\mu \tilde A^b_\nu \rightarrow h_{\mu \nu}$,
dropping the color factors.  (The color indices play essentially no
role in this identification.)

If one had an all orders Lagrangian, it would be possible to
investigate non-perturbative implications.  Because of the double-copy
property, one might expect that {\it all} classical solutions of
gravity could be written as double copies of solutions of gauge
theories.  In particular, in coordinate space we can expect
gravity solutions to be convolutions of gauge-theory
solutions, with the schematic
form,
$
g_{\mu\nu}(x) \sim \int d^Dy \, A_\mu^a(x-y) \tilde A^b_\nu(y).
$

\section{The color-kinematic duality at loop level}

Very recently ref.~\cite{BCJLoops} proposed that the color-kinematic
duality and gravity double-copy property extends to {\it all} loop
orders.  This might seem a rather bold conjecture, yet the unitarity
method provides strong motivation.  Indeed the duality was first
observed in the three- and four-loop four-point amplitudes of
$\NeqFour$ super-Yang-Mills theory, with various on-shell conditions
imposed~\cite{BCJ}.  The success of the duality at the Lagrangian
level~\cite{Square} is also quite suggestive.

The loop-level expressions for gauge and gravity amplitudes would then be, 
\begin{eqnarray}
  (-i)^L {\cal A}^\Loop_n \hskip -.2 cm \!\!  &=& \! \hskip -.2 cm 
 \! \sum_{j}{\int{\prod_{l = 1}^L {d^D p_l \over (2 \pi)^D}
  \frac{1}{S_j}  
 \frac {n_j c_j}{\prod_{\alpha_j}{p^2_{\alpha_j}}}}}\,, \label{LoopBCJ}
\\
 (-i)^{L+1}  {\cal M}^\Loop_n \hskip -.2 cm \!\! &=& \! \hskip -.2 cm 
 \sum_{j} {\int{ \prod_{l = 1}^L {d^D p_l \over (2 \pi)^D}
 \frac{1}{S_j}
   \frac{n_j \n_j}{\prod_{\alpha_j}{p^2_{\alpha_j}}}}} \,, \nn
\end{eqnarray}
where we again suppressed the couplings and the sums now run over all
distinct $n$-point $L$-loop diagrams with cubic vertices.  These
include distinct permutations of external legs, and the $S_j$ are the
symmetry factors of each diagram.  As at tree level, at
least one family of numerators ($n_j$ or $\n_j$) is constrained to
satisfy the duality~(\ref{BCJDuality}).  We expect these formulas to
hold in a large class of theories, including theories which are the
low energy limits of string theories.  It should also hold in
pure gravity, but in this case extra projectors would be
required to remove the extra unwanted states arising in the direct product
of two pure Yang-Mills theories.

As at tree level, the ability to organize amplitudes around cubic graphs is
trivially accomplished by inserting inverse propagators into the 
numerators to account for
contact terms.  The non-trivial part of this conjecture
is that there always exists sufficient freedom to arrange gauge-theory
multiloop amplitudes in a way that satisfies the color-kinematic
duality (\ref{BCJDuality}).  The unitarity method straightforwardly
ensures that the double-copy property of gravity extends to loop level
when it holds at tree level, since all cuts that decompose loop
amplitudes into products of tree amplitudes will have this
property~\cite{BCJLoops,Square}.

\subsection{Checks on the loop-level conjecture}

Since we do not yet have a proof that the color-kinematic duality
holds, it is important to check some explicit examples.  The known one
and two-loop four-point amplitudes of $\NeqFour$ super-Yang-Mills
theory and $\NeqEight$ supergravity, as given in ref.~\cite{BDDPR},
satisfy the conjectures (\ref{LoopBCJ}).  Another example is the
previously mentioned two-loop four-point identical-helicity amplitude
of pure Yang-Mills theory~\cite{TwoLoopAllPlus}, which can also be
shown to satisfy the duality~\cite{BCJ,BCJLoops}.

A more sophisticated example, which we outline here, is the three-loop
four-point amplitude of $\NeqFour$ super-Yang-Mills theory.  This
amplitude offers a rather non-trivial check of the duality given both
the high loop order and the non-trivial dependence of the numerators
on loop momenta. It was originally constructed in
refs.~\cite{GravityThree,ManifestThreeLoop} via the unitarity method.
In ref.~\cite{BCJLoops}, the duality was made manifest by appropriate
reshufflings of terms in the earlier forms of the amplitude.

\begin{figure}[tb]
\begin{center}
\includegraphics[clip,scale=0.4]{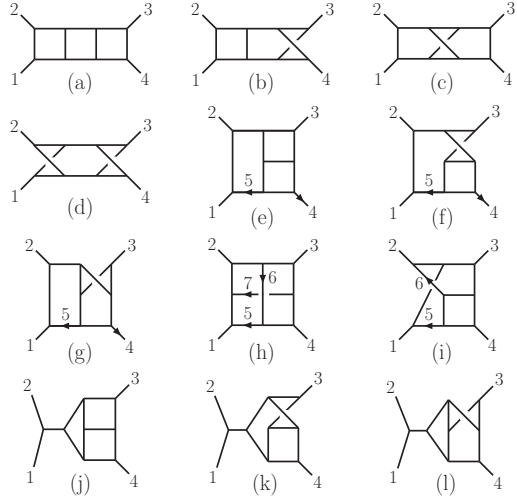}
\end{center}
\vskip -1. cm 
\caption[a]{\small The 12 diagrams contributing to both $\NeqFour$
  super-Yang-Mills and $\NeqEight$ supergravity three-loop four-point
  amplitudes in the arrangement of ref.~\cite{BCJLoops}. The
  corresponding integrals are obtained by combining their propagators
  with numerator factors given in \tab{NumeratorTable}. The (internal)
  symmetry factor for diagram (d) is $S_{\rm (d)}=2$, the rest are
  unity.  All distinct external permutations of each diagram appear in
  the amplitude.}
\label{DiagramsFigure}
\end{figure}

\begin{figure}[tb]
\vskip -.5 cm 
\begin{center}
\includegraphics[clip,scale=0.56]{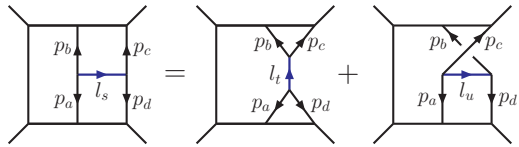}
\end{center}
\vskip -1. cm 
\caption[a]{\small A numerator duality relation at three loops.
  }
\label{loopJacob}
\end{figure}

The duality requires that the numerator identities in \eqn{BCJDuality}
must be imposed for every propagator in every diagram.  In
\fig{loopJacob} we display one such numerator relation.  For any
diagram, we can describe any internal line, carrying some momentum
$l_s$, in terms of formal graph vertices $V(p_a,p_b,l_s)$, and
$V(-l_s, p_c, p_d)$ where the $p_i$ are the momenta of the other legs
attached to $l_s$, as illustrated on the left side of \fig{loopJacob}.
The duality (\ref{BCJDuality}) requires that,
\begin{eqnarray}
&& \hskip -.8 cm 
n(\{\,V(p_a, p_b, l_s),\,V(-l_s, p_c, p_d),\,\cdots\,\})= \nn\\
&& \null \hskip .3 cm 
 n(\{\,V(p_d,p_a,l_t),\,V(-l_t,p_b,p_c),\,\cdots\,\}) \nn \\
&& \null \hskip .3 cm 
+n(\{\,V(p_a,p_c,l_u),\,V(-l_u,p_b,p_d),\,\cdots\,\})\, ,  \nn
\end{eqnarray}
where $n$ represents the numerator associated with the diagram
specified by the set of vertices, the omitted vertices are identical
in all three diagrams, and $l_s \equiv \left(p_c+p_d\right)$, $l_t
\equiv\left(p_b+p_c\right)$ and $l_u\equiv\left(p_b+p_d \right)$ in
the numerator expressions.  There is one such equation for every
propagator in every diagram.  Solving the system of distinct equations
enforces the duality conditions~(\ref{BCJDuality}).  It turns
out\cite{BCJLoops} that a solution to the duality relations is found
in terms of the 12 diagrams displayed in \fig{DiagramsFigure}.

\begin{table*}
\caption{The numerator factors of the integrals in
\fig{DiagramsFigure}. The first column labels the integral, the second
column the relative numerator factor for $\NeqFour$ super-Yang-Mills
theory.  The square of this is the relative numerator factor for
$\NeqEight$ supergravity. 
\label{NumeratorTable} }
\vskip .4 cm
\begin{center}
\begin{tabular}{||c|c||}
\hline
Integral $I^{(x)}$ &  $\NeqFour$ Super-Yang-Mills ($\sqrt{\NeqEight~{\rm supergravity}}$) numerator  \\
\hline
\hline
(a)--(d) &  $s^2$   \\
\hline 
(e)--(g) & $[s \left(-\tau _ {3 5}+\tau _ {4 5}+t \right)- t \left(\tau _ {2 5}+\tau _ {4 5}\right)+
 u \left(\tau _ {2 5}+\tau _ {3 5}\right)-s^2 ]/3$   \\
\hline
(h)& $ [s \left(2 \tau _ {1 5}-\tau _ {1 6}+2 \tau _ {2 6}-\tau _ {2 7}+2 \tau _ {3 5}+\tau _ {3 6}+\tau _ {3 7}-u \right)$\\
&$\null + t \left(\tau _ {1 6}+\tau _ {2 6}-\tau _ {3 7}+2\tau _ {3 6}-2 \tau _ {1 5}-2\tau _ {2 7}-2\tau _ {3 5}-3 \tau _ {1 7}\right)+s^2 ]/3$\\
\hline
(i)& $[s \left(-\tau _ {2 5}-\tau _ {2 6}-\tau _ {3 5}+\tau _ {3 6}+\tau _ {4 5}+2 t \right)$\\
&$\null 
+ t \left(\tau _ {2 6}+\tau _ {3 5}+ 2\tau _ {3 6}+2\tau _ {4 5}+3 \tau _ {4 6}\right)+ u\,\tau _ {2 5}+s^2 ]/3$\\
\hline
(j)-(l) & $s (t-u)/3 $
 \\
\hline
\end{tabular}
\end{center}
\end{table*}

Of course the amplitude must be the correct one for the theory under
consideration, and thus must satisfy all cut conditions as well.  The
explicit solution found in ref.~\cite{BCJLoops}, satisfying both the
cut conditions and the duality, is displayed in \tab{NumeratorTable}.
Interestingly, after imposing the duality, the maximal cut of diagram
(e) in \fig{DiagramsFigure} is sufficient for finding the unique
solution, when the numerators are local and no diagram has a
worse power counting than the known amplitude.  In this table, an
overall factor of $s t A_4^\tree$ has been removed from the entries,
$s,t,u$ are Mandelstam invariants corresponding to $(k_1 + k_2)^2,(k_2
+ k_3)^2, (k_1 + k_3)^2$ and $\tau_{i j} = 2 k_i\cdot l_j$, where
$k_i$ and $l_j$ are external and internal momenta
as labeled in \fig{DiagramsFigure}.  The
reader may check that all duality relations hold.
Ref.~\cite{BCJLoops} also verified that the double-copy property holds
for $\NeqEight$ supergravity, confirming it on a complete set of cuts
using the known amplitude~\cite{GravityThree,ManifestThreeLoop}.

\section{Comments and conclusions}

In this talk we summarized a recently proposed gauge-theory duality
between color and kinematics, leading to a double-copy property for
gravity theories~\cite{BCJ,BCJLoops}.  Although the duality remains a
conjecture, we can even now exploit it to guide loop computations,
simply by enforcing the duality and verifying the consistency with the
unitarity cuts.  This is helpful both for organizing gauge-theory
amplitudes and for obtaining the corresponding double-copy
gravity amplitudes.

There are a number of interesting open problems related to the
color-kinematic duality.  In particular, it would be helpful to carry
out further checks of the duality for multiloop processes.  More
generally an all-orders proof would be important, especially if it
leads to new insight into the origins of the duality. The duality
suggests the existence of a group theoretical construction of the
kinematic numerators, which would, of course, be very interesting to
develop. We would also like to have Lagrangians whose diagrams satisfy
the duality to all orders, and which give gravity Lagrangians as
double copies, along the lines described in ref.~\cite{Square}.  We
expect this to have non-trivial implications at strong coupling.  It
seems likely that the duality should hold as well in higher-genus
perturbative string theory.

The duality and double-copy property may also shed new light on the
issue of ultraviolet divergences in $\NeqEight$ supergravity.  For four-point
amplitudes through four loops, explicit computations show that ultraviolet
cancellations exist beyond those needed for
finiteness~\cite{BDDPR,GravityThree,ManifestThreeLoop,GravityFour}. Beyond
this, the situation is less clear.  A consensus has formed that
supersymmetry alone cannot prevent divergences in four dimensions
starting at seven
loops~\cite{KellyYM,SevenLoopDiv,NineLoopRetraction}.\footnote{The
  claimed delay from supersymmetry of potential ultraviolet divergences in
  $\NeqEight$ supergravity until nine loops~\cite{NineLoopWrong} has
  been retracted~\cite{NineLoopRetraction}.} Although nontrivial
cancellations are known to exist to all loop orders~\cite{Finite}, we
do not know if these cancellations are sufficient to render the theory
finite.  Recent reviews discussing the ultraviolet properties of
$\NeqEight$ supergravity may be found in refs.~\cite{GravityUVReview}.

The color-kinematic duality and gravity double-copy structure likely
have important non-perturbative implications.  In particular, these
properties suggest that classical solutions in gravity theories may be
expressible as double copies of classical solutions in gauge theories.

\vskip .3 cm 

We thank T.~Dennen, L.~Dixon, Y.~t.~Huang, H.~Ita, M.~Kiermaier and
R.~Roiban for discussions and collaboration on topics described in
this presentation.  This research was supported by the US Department
of Energy under contract DE--FG03--91ER40662. J.J.M.C. gratefully
acknowledges the financial support of a Guy Weyl Physics and Astronomy
Alumni Grant.  H.J.'s research is supported by the European Research
Council under Advanced Investigator Grant ERC-AdG-228301.


\end{document}